\documentclass[runningheads]{llncs}
\usepackage{graphicx}
\usepackage{booktabs}
\usepackage{amsmath}
\usepackage{amsfonts}
\usepackage{multirow}
\usepackage{tabularx}
\newcolumntype{C}[1]{>{\centering\arraybackslash}p{#1}}
\usepackage[T1]{fontenc}

\usepackage{threeparttable}
\usepackage{color}
\usepackage[normalem]{ulem}
\usepackage{array}
\usepackage[table,dvipsnames]{xcolor}
\usepackage{colortbl}
\usepackage{wrapfig}
\usepackage{lipsum}

\usepackage{adjustbox}
\usepackage{subfigure}
\usepackage{paralist} 
\usepackage{enumitem} 
\usepackage{booktabs} 
\usepackage{array}

\usepackage{xspace}
\graphicspath{{figures/}}

\usepackage[pagebackref=true,breaklinks=true,letterpaper=true,colorlinks,citecolor=NavyBlue,linkcolor=BrickRed,bookmarks=false]{hyperref}

\long\def\ignorethis#1{}

\definecolor{gray}{rgb}{0.35,0.35,0.35}
\definecolor{MyBlue}{rgb}{0,0.2,0.8}
\definecolor{MyRed}{rgb}{0.8,0.2,0}
\definecolor{MyGreen}{rgb}{0.0,0.5,0.1}
\definecolor{MyGray}{rgb}{0.4,0.4,0.4}

\newlength\paramargin
\newlength\figmargin
\newlength\secmargin

\newcolumntype{L}[1]{>{\raggedright\let\newline\\\arraybackslash\hspace{0pt}}m{#1}}
\newcolumntype{C}[1]{>{\centering\let\newline\\\arraybackslash\hspace{0pt}}m{#1}}
\newcolumntype{R}[1]{>{\raggedleft\let\newline\\\arraybackslash\hspace{0pt}}m{#1}}

\def\ie{i.e.,~}

\newcommand{\xx}{\mathbf{x}}
\newcommand{\mm}{\mathbf{m}}
\newcommand{\uu}{\mathbf{u}}
\newcommand{\DV}{\mathcal{D}^\text{v}}
\newcommand{\DC}{\mathcal{D}^\text{c}}
\newcommand{\DHH}{\mathcal{D}^\text{h}}
\newcommand{\dd}{\mathbf{d}}
\newcommand{\FQ}{Q}
\newcommand{\FE}{E}
\newcommand{\FG}{D}
\newcommand{\argmin}{\operatornamewithlimits{argmin}}

\newcommand{\zz}{\mathbf{z}}
\usepackage{mathtools}
\newcommand{\sg}{\operatorname{sg}}

\usepackage{graphicx}
\def\ourmdoel{HiCo-Net}

\usepackage[capitalize]{cleveref}
\crefname{section}{Sec.}{Secs.}
\Crefname{section}{Section}{Sections}
\Crefname{table}{Table}{Tables}
\crefname{table}{Tab.}{Tabs.}
\begin{document}
\title{Collaborative Quantization Embeddings for Intra-Subject Prostate MR Image Registration}
\titlerunning{\ourmdoel}
%
%
\authorrunning{Z. Shen et al.}

\author{Ziyi Shen\inst{1}\and
Qianye Yang\inst{1}\and
Yuming Shen\inst{2}\and
Francesco Giganti\inst{3}\and
Vasilis Stavrinides\inst{1}\and
Richard Fan\inst{4}\and
Caroline Moore\inst{1}\and
Mirabela Rusu\inst{4}\and
Geoffrey Sonn\inst{4}\and
Philip Torr\inst{2}\and
Dean Barratt\inst{1}\and
Yipeng Hu\inst{1}
}

\institute{University College London, London, UK \and
University of Oxford, Oxford, UK \and
 University College London Hospital NHS Foundation
Trust, London, UK \and
Stanford University, CA 94305, USA\\
\email{joanshen0508@gmail.com}
}
%
\maketitle              
\begin{abstract}

%
%
Image registration is useful for quantifying morphological changes in longitudinal MR images from prostate cancer patients. 
This paper describes a development in improving the learning-based registration algorithms, for this challenging clinical application often with highly variable yet limited training data. 
First, we report that the latent space can be clustered into a much lower dimensional space than that commonly found as bottleneck features at the deep layer of a trained registration network. Based on this observation, we propose a hierarchical quantization method, discretizing the learned feature vectors using a jointly-trained dictionary with a constrained size, 
in order to improve the generalisation of the registration networks. Furthermore, a novel collaborative dictionary is independently optimised to incorporate additional prior information, such as the segmentation of the gland or other regions of interest, in the latent quantized space. Based on 216 real clinical images from 86 prostate cancer patients, we show the efficacy of both the designed components. 
Improved registration accuracy was obtained with statistical significance, in terms of both Dice on gland and target registration error on corresponding landmarks, the latter of which achieved 5.46 mm, an improvement of 28.7\% from the baseline without quantization. Experimental results also show that the difference in performance was indeed minimised between training and testing data. 

%
%
%
%
%
%
%

\end{abstract}
\section{Introduction}\label{sec_1}
%
Whilst the diagnostic value in multiparametric MR imaging for prostate cancer, before or after biopsy for histopathology examination, has been identified \cite{moore2017reporting,schoots2015magnetic}, attention has been quickly turned to using this non-invasive imaging technique to monitor the disease at its early stage. 
Many have speculated that the temporal changes in morphology and intensity pattern in prostate gland can indicate the progression of the cancer \cite{kim2008mri}. 
Establishing spatial correspondence between two or more images, medical image registration has been proposed for aligning MR images from prostate cancer patients acquired at different time points. The estimated intra-subject spatial transformation, representing corresponding spatial locations between prostate glands, is an important quantitative tool to track the radiological evolution of prostate glands~\cite{yang2020longitudinal}. 
Aligning anatomical structures in lower pelvic region has been recognised to be challening using `classical' iterative algorithms \cite{zhang2017frequency}, perhaps due to the highly patient-specific imaging characteristics in organs, such as prostate glands being having distinct patient-specific intensity patterns on T2-weighted MR images. Recent development has focused on learning-based algorithms for their effective and efficient inference \cite{affine_chen2021learning,2019voxelmorph,mok2020large,wang2020deepflash,kim2021cyclemorph,xu2021multi}. 
Empirically, only a few `types' of MR image features are reliably useful for establishing correspondence, such as volume, shape, anatomical and pathological structures, within or around the prostate gland and potential cancerous regions \cite{yang2020longitudinal}. 
In addition, prostate capsules and different types of pathology are known to be highly variable and specific to individual patients \cite{song2021cross,zhang2021learning}. 
Therefore, features are more likely to be `easy-to-learn' to the distinct inter-subject differences.

We argue that these two above intuitions may 
warrant a compact feature adequately containing intra-subject correspondence for this application, although a deep network may still be required to learn such a representation \cite{liu2021same}.
However, deep models are over-parametrized~\cite{2017variational}, where the hidden representation 
may still carry information that is not related the task.
With limited MR training data in a real clinical application, this redundancy and over-parametrization degrade the learned features' ability to generalize, leading to overfitting.

\begin{figure}[t]
	\begin{center}
		\includegraphics[width=\textwidth]{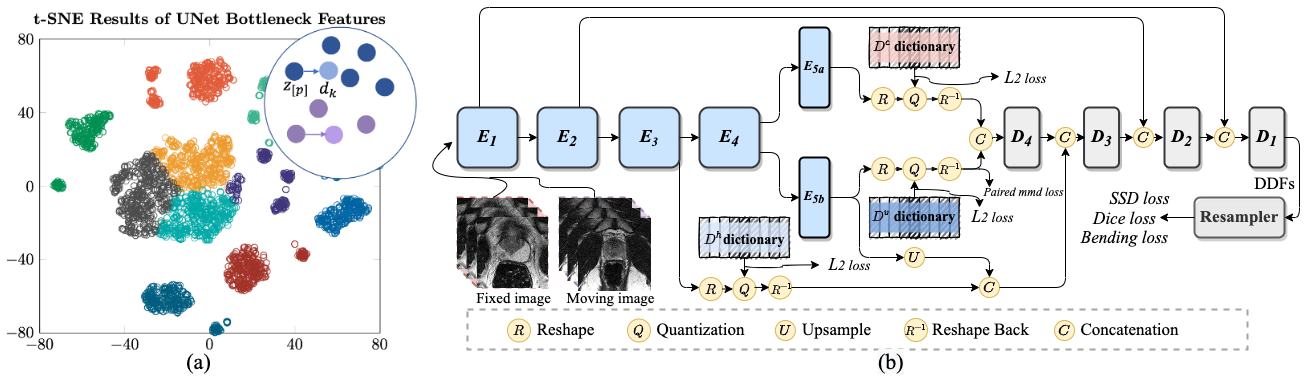}
	\end{center}
\caption{Demonstration of the proposed prostate registration framework. (a)~The t-SNE \cite{tsne} visualization of the encoder outputs of a U-Net-like prostate registration network. (b)~The network structure of \ourmdoel.}
\label{fig:framework}
\end{figure}
To illustrate this redundancy, we visualize the t-SNE \cite{tsne} results of the deep features for registering prostate MR images in \cref{fig:framework}~(a).
Given an $N$-sample prostate image set, we extract the bottleneck features of all samples, producing $N$ feature maps with a size of $W\times H\times T\times C$. 
Here, $W\times H\times T$ refers to the shape of the output feature, while $C$ is the channel number. 
Each $C$-dimensional vector represents a super-pixel in the encoded feature. 
We visualize the t-SNE result of all super-pixels in the set. 
It is shown in \cref{fig:framework}~(a) that the features are scattered into a limited numbers of groups.
In other words, one can roughly represent the whole feature space using a smaller number of latent topics, consistent with the above discussion in limited corresponding structures and inter-subject variability.
%

%
This observation opens the door to compress the features with deep vector quantization (VQ) \cite{peng2021generating,razavi2019generating,van2017neural}.
We refer to the anatomical knowledge of prostate, whose appearance varies from subjects, and propose to represent the neural features with a small set of vocabulary vectors, where useful anatomical and pathological structures are preserved.    
%
%
%
This ideology has been proved effective with limited training samples \cite{chen2020incremental}, which is indeed the case our task.

In this paper, the aforementioned motivation drives us to a VQ-based prostate registration solution. 
Specifically, we apply VQ to the middle of a registration U-Net \cite{unet} to make effective use of the feature space. 
The dictionary of a vector quantizer is usually data-driven. 
To efficiently improve the anatomical and pathological awareness of the model, we introduce another quantizer in parallel to the randomly initialized one, of which the dictionary preserves the specific local features of prostate interior.
This is done by abstracting knowledge from a deep prostate segmentation network. 
The combination of the data-driven and the pre-defined quantizers is termed \textbf{collaborative quantization}. In addition, we  explore the multi-scale structure of prostate MR images to fit both the global and local patterns of a moving image to the fixed one, which suggests a \textbf{hierarchical quantization} structure similar to \cite{razavi2019generating}.
Therefore, we name our model as Hierarchically \& Collaboratively Quantized Network (\ourmdoel). 

Our contributions can be summarised as follows:~(1) We propose a feature quantization framework as regularization to alleviate the gap between training and test data for registration~(2) We introduce a collaborative quantizer that encodes structure features of the gland boundary to better represent lesions and landmarks.~(3)~Our experiments show that the proposed \ourmdoel successfully relieves the overfitting problem in weakly-supervised longitudinal prostate image registration, and outperforms the state-of-the-art.

\section{Method}

We consider a pairwise MR image registration problem. Let $\mathcal{X}=\{\xx_i\}_{i=1}^n$ be the collection of images pairs of prostate, where $\xx_i = (\xx_i^\text{s}, \xx_i^\text{t})$ denotes a pair, with $n$ being the number of image pairs. $\xx_i^\text{s}$ and $\xx_i^\text{t}$ respectively refers to the moving image and the fixed one. Each image pair comes with a pair of prostate gland anatomical segmentation maps $(\mm_i^\text{s},\mm_i^\text{t})$ for weakly-supervised training. For each $\xx_i$, the goal is to predict a dense displacement field (DDF) $\uu_i$ to establish voxel-level correspondence.

%

\subsection{Preliminary: Deep Vector Quantization}
VQ \cite{van2017neural} quantizes an arbitrary representation tensor using a fixed number of values defined by a dictionary $\mathcal{D}=\{\dd_i\}_{i=1}^K,~\dd_i\in\mathbb{R}^C$, where $K$ is the dictionary size and $C$ is the dimensionality of each code. Specifically, an output of an encoder~$E(\xx) \in \mathbb{R}^{H \times W \times T \times C}$ is obtained by passing an MR image pair $\xx$ through a CNN. We are going to quantize each $C$-dimensional vector of $E(\xx)$. For simplicity, the rest of this paper denotes a voxel  super-position in a raster scan order with a coordinate $p=1\cdots H WT$, \ie $E(\xx)_{[p]}\in\mathbb{R}^C$. Hence, a vector quantization operator $\FQ(\cdot)$ can be defined as follows:
\begin{equation}\label{eq_1}
    \zz_{[p]}=\FQ\left(\FE(\xx)_{[p]}; \mathcal{D}\right) = \dd_k,~\text{where}~k=\argmin_i\|\FE(\xx)_{[p]}-\dd_i\|.
\end{equation}
Then, $\zz$ replaces $\FE(\xx)$ and is forwarded to the rest of the network. The encoder receives the gradients from top of the quantizer through the straight-through estimator~\cite{ste}, \ie $\partial\zz/\partial\FE \coloneqq \mathbb{I}$. VQ incorporates two additional loss terms to enforce the output of the encoder to be similar to the quantized results:
\begin{equation}
    \mathcal{L}_\text{Q}(\FE(\xx),\mathcal{D}) = \sum_p\left\|\sg\left(\FE(\xx)_{[p]}\right)-\zz_{[p]}\right\|^2_2 +\beta\left\|\FE(\xx)_{[p]}-\sg\left(\zz_{[p]}\right)\right\|_2^2,
\end{equation}
where $\sg(\cdot)$ is the stop-gradient operator and $\beta=0.25$ is the hyperparameter.

\subsection{Model Overview}
\cref{fig:framework}~(b) depicts the schematic of \ourmdoel. It generally undergoes a U-Net-like structure with an encoder $\FE(\cdot)$ and a decoder $\FG(\cdot)$. An image pair $\xx=(\xx^\text{s},\xx^\text{t})$ is firstly concatenated together and then rendered to the encoder, while the decoder produces $\uu$, the desired DDF. We particularly denote the output of each residual block as $\FE_1(\xx), \FE_2(\xx)\cdots$. Notably, $\FE_4(\xx)$ is rendered to two parallel convolutional layers, $\FE_{5\text{a}}(\xx)$ and $\FE_{5\text{b}}(\xx)$, that are followed be two quantizers. We respectively term them the collaborative quantizer (\cref{sec_24}) and the vanilla one (\cref{sec_23}). The skip connection between $\FE_3(\xx)$ and the decoder is quantized as well. Since it also mixes multi-scale information from $\FE_{5\text{b}}(\xx)$ afterwards, we name it the hierarchical quantizer (\cref{sec_25}). The intuition behind this design is given in their respective sections as follows. The DDF output then contributes to the conventional weakly-supervised prostate registration losses with a resampler.

\subsection{Vanilla Quantization}\label{sec_23}
We first introduce a vanilla quantizer that quantizes the output of $\FE_{5\text{b}}(\xx)$. It behaves identical to the original VQ operation \cite{van2017neural}. Shown in \cref{fig:framework}~(b), its dictionary $\DV$ is randomly initialized and is updated by back-propagation during training. In this way, the global information, a relatively fixed structure of the MR prostate images, is regularized for better generalization to test data. We denote the quantization loss for the vanilla quantizer as $\mathcal{L}_\text{V}(\xx) = \mathcal{L}_\text{Q}(\FE_{5\text{b}}(\xx), \DV)$.

\subsection{Hierarchical Quantization}\label{sec_25}
Image features of a deep network often carry local information, which can benefit from multi-scale modelling for positional alignment. 
A hierarchical representation quantizer has been proved to be effective to perceiving this \cite{razavi2019generating}.

To implement this, we employ a hierarchical quantizer to quantize the output of $\FE_3(\xx)$, of which the dictionary is denoted as $\DHH$. The quantized result is added by the output of $\FE_4(\xx)$. Since the voxel sizes of them mismatch, one needs to firstly upsample $\FE_4(\xx)$, as is shown in \cref{fig:framework}~(b). 
$\DHH$ is randomly initialized. The hierarchical quantizer introduces a quantization loss as $\mathcal{L}_\text{H}(\xx) = \mathcal{L}_\text{Q}(\FE_{3}(\xx), \DHH)$.


\subsection{Collaborative Quantization}\label{sec_24}

As is discussed in \cref{sec_1}, the awareness of prostate contour is of key importance to this prostate registration task. This inspires us to transfer knowledge to our model from a deep segmentation network, without requiring segmentation during inference. 
VQ allows us to conveniently initialize the dictionary values according to the segmentation network's output as prior knowledge. In particular, we first train a U-Net-based segmentation network on the training dataset, and then extract an $H\times W\times T\times C$ tensor for each image with its encoder. 
Each $C$-dimensional vector of all images' features is treated as an instance for K-means clustering. We cache the values of $K$ cluster centers produced by K-means, and use them to initialize the dictionary of the collaborative quantizer, \ie $\DC$. An analogy of this procedure is provided in the supplemental. 

The collaborative quantizer takes input from $\FE_{5\text{a}}(\xx)$ as a compensation to the fully-data-driven vanilla quantizer by concatenating their outputs afterwards. We similarly define its quantization loss as $\mathcal{L}_\text{C}(\xx) = \mathcal{L}_\text{Q}(\FE_{5\text{a}}(\xx), \DC)$.

\subsection{Training}\label{sec_26}
\noindent\textbf{Quantization Loss.} The training objective of \ourmdoel~is a combination of the quantization losses and the conventional weakly-supervised registration ones. We first define the overall quantization loss of an image pair $\xx$:
\begin{equation}
    \mathcal{L}_\text{Quant}(\xx) = \mathcal{L}_\text{V}(\xx) + \mathcal{L}_\text{C}(\xx) +\mathcal{L}_\text{H}(\xx).
\end{equation}
The three quantization terms above can have equal weights as they are mutually independent, and are imposed to different stages of the registration network.

\noindent\textbf{SSD Loss.} The Sum-of-Square Differences (SSD) loss \cite{2019voxelmorph} measures the similarity between the translated image and the fixed one. One needs to firstly resample $\xx^\text{s}$ using the DDF $\uu$ and then compute
\begin{equation}
    \mathcal{L}_\text{SSD}(\xx) =\left\|\uu\otimes\xx^\text{s} - \xx^\text{t}\right\|_2^2,
\end{equation}
where $\otimes$ refers to the resampling operation.

\noindent\textbf{Dice Loss.} This loss has shown effectiveness in aligning organ shapes and positions \cite{xu2019deepatlas}, and is applied to the masks:
\begin{equation}
    \mathcal{L}_\text{Dice}(\xx) =-\operatorname{Dice}\left(\uu\otimes\mm^\text{s},~ \mm^\text{t}\right).
\end{equation}

\noindent\textbf{Bending Regularization.} We use this regularization term $\mathcal{L}_\text{Bend}(\uu)$ \cite{zeng2020label} to penalise the non-smoothness of the generated DDF.

\noindent\textbf{Overall Training Objective.} By defining the losses above, we can simply compose a linear combination of them as the final loss of \ourmdoel~as follows:
\begin{equation}
    \mathcal{L}_\text{All}(\xx) = \lambda_\text{Q}\mathcal{L}_\text{Quant}(\xx)+ \lambda_\text{S}\mathcal{L}_\text{SSD}(\xx) + \lambda_\text{D}\mathcal{L}_\text{Dice}(\xx) + \lambda_\text{B}\mathcal{L}_\text{bend}(\uu),
\end{equation}
where $\lambda_\text{Q}=1$, $\lambda_\text{S}=1$, $\lambda_\text{D}=1$ and $\lambda_\text{B}=50$ are hyperparameters. Our models are trained with stochastic gradient descent algorithms.
\begin{figure}[t]
\centering
\begin{minipage}{.49\textwidth}
  \centering
  \includegraphics[width=0.9\linewidth]{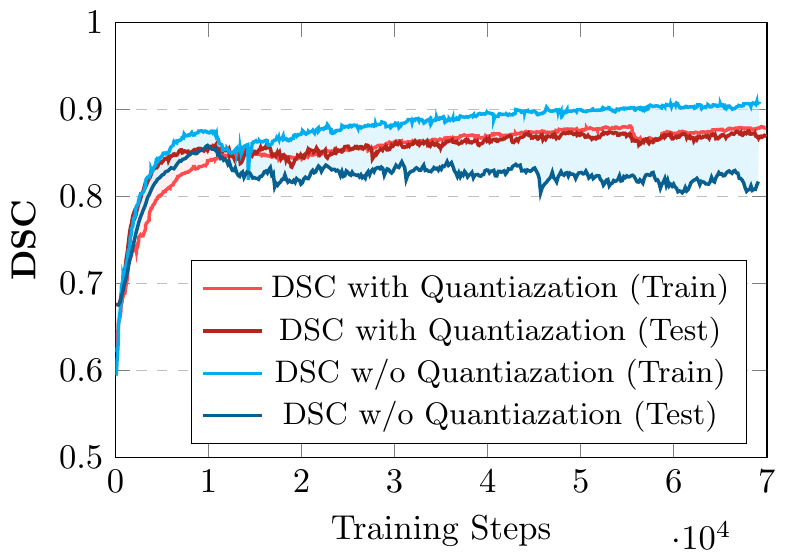}
  \caption{Illustration of how $\DC$ is initialized before training.}
  \label{fig:train}
\end{minipage}\hfill
\begin{minipage}{.45\textwidth}
  \centering
  \includegraphics[width=1.0\linewidth]{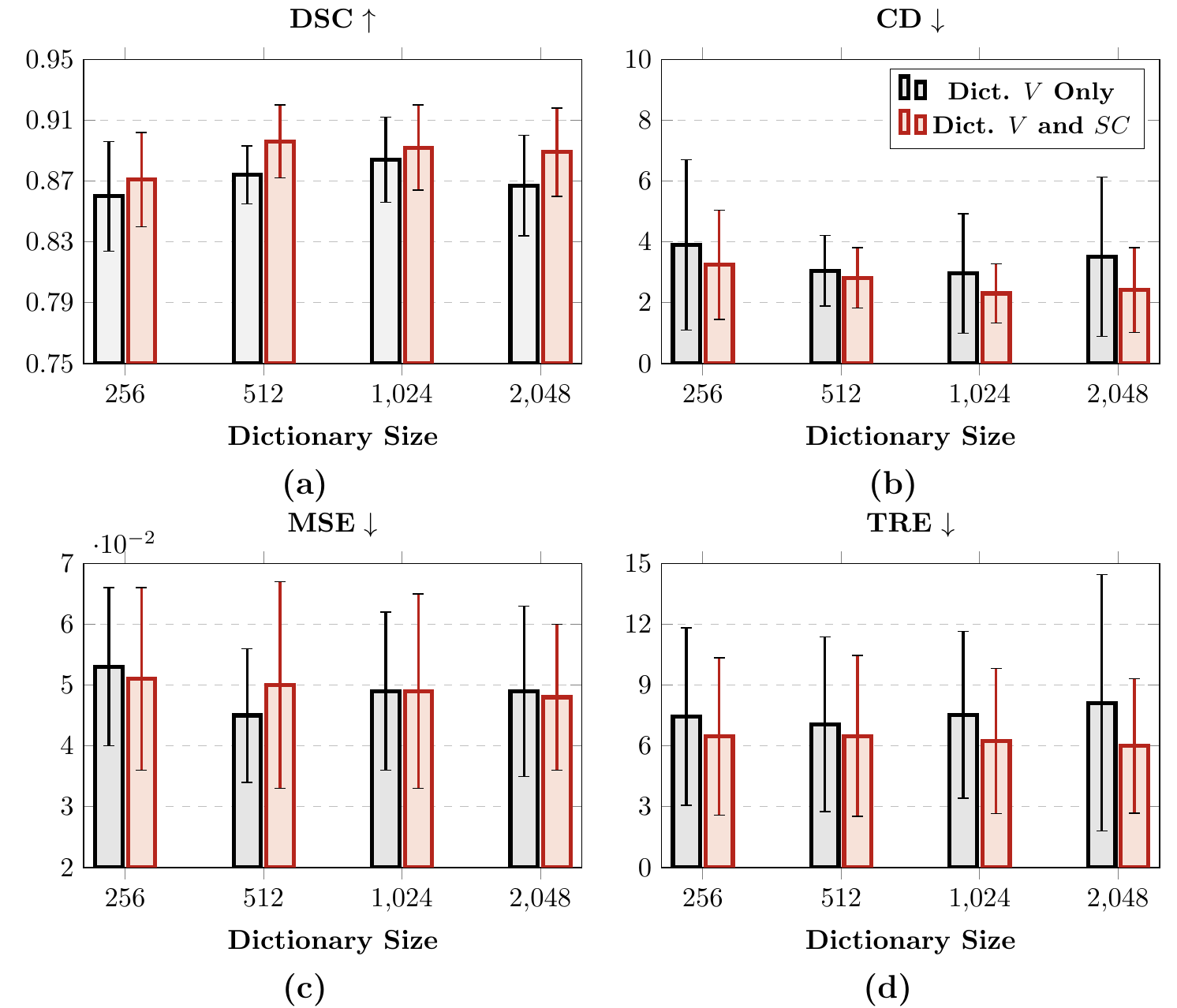}
  \caption{Ablation study of dictionary size.}
  \label{fig:hp}
\end{minipage}
\end{figure}
\section{Experiment}
\subsection{Experimental Settings}
\noindent\textbf{Implementation.} We use a basic U-Net equipped with skip connections between encoder and decoder.
Our encoder consists of 4 residual blocks, a total of 12 convolutional layers.
%
%
In addition, we add 2 convolutional layers to expand heads for the subsequent hierarchical and collaborative quantization operators as shown in~\cref{fig:framework}~(b).
%
The channel size of the hierarchical dictionary $\DHH$ is 128.
The vanilla dictionary $\DV$ and collaborative one $\DC$ both have a feature channel number of 256.
We by default set the vocabulary size of $\DHH$ and $\DV$ to 1024, while fixing the size of $\DC$ one to 512.
%
%
%
As per the initialization of $\DC$, we use a U-Net as our segmentation network, which is utilized to collect features for K-means clustering.
Note that the segmentation network is not involved in our registration training. We set training batch size to 4, and use the Adam optimizer with a learning rate of $10^{-4}$. %
The network is trained for 1000 epochs at most, taking three days on an NVIDIA Tesla V100 GPU.
%
%
The network architecture and code are available at \url{https://github.com/joanshen0508/HiCo-Net}. 

%
%

%

\noindent{\textbf{Dataset.}} 
The utilized dataset consists of 216 longitudinal prostate T2-weighted MR images from 86 patients, acquired from University College London Hospitals NHS Foundation.
It is divided into three folds, containing 70, 6, and 10 patients for training, validation, and test. Each patient has 2-4 images, with an average interval between consecutive visits of 18.1 and a standard deviation of 10.3 months.
Before training, we resample the data to $0.7\times0.7\times0.7 mm^3$ and normalize the intensity to $[0,1]$.
To train the proposed prostate registration model, we also crop the dataset 
and generate the dataset with the size of $128\times128\times102$. 
On the test set, 141 anatomical and pathological landmarks are manually identified on moving and fixed images, including patient-specific fluid-filled cysts, calcification and centroids of zonal boundaries.

\noindent\textbf{Evaluation Metrics.} We adopt the conventional weakly-supervised registration metrics including Dice Similarity Coefficient~(DSC) and Centroid Distance (CD) between the prostate glands. The Mean-Squared Error (MSE) between the fixed image and wrapped moving image is as well reported.
Registration should support downstream clinical image analysis task. To demonstrate the effectiveness of our method, we further report the Target Registration Error(TRE), which calculates the difference of landmarks between fixed image and predicted result.

\subsection{Ablation study}
\subsubsection{The Effect of Feature Quantization.} 
%
We first build a registration model without quantization similar to \cite{yang2020longitudinal} and compare it with a variant of \ourmdoel~that only involves the vanilla quantizer. 
As shown in~\cref{fig:train}, 
our quantized version effectively narrows the accuracy gap between training and test, observing no overfitting problem.
%
%
We also report the performance of \ourmdoel~with different combinations of the three quantizers in~\cref{ablation:size}. 
%
Compared with the unquantized baseline, the TRE is reduced from $7.657\pm4.212$ mm to $6.248\pm3.577$ mm ($\mathtt{p\_value}=0.0001$ under paired t-test) when applying the vanilla and collaborative quantization, and it further decreases to $5.457\pm 3.489$ mm ($\mathtt{p\_value}<0.0001$) when employing all the three quantizers.
%

\begin{table}[t]
\centering
	\caption{Ablation study of hierarchical and collaborative quantization. }\label{ablation:size}
	\begin{tabular}{p{1.0cm}p{1.0cm}l|c|c|c|c}
		\hline
		$\DV$&~$\DHH$&~$\DC$&\textbf{DSC}&\textbf{CD}&\textbf{MSE}&\textbf{TRE}\\
		\hline
		\multicolumn{3}{l|}{w/o registration}&0.700$\pm$0.097&12.63$\pm$5.810&0.051$\pm$0.014&13.72$\pm$5.833\\
		\hline
		&&&0.859$\pm$0.038&4.187$\pm$2.050&0.049$\pm$0.013&7.657$\pm$4.212\\
		\checkmark&&&0.884$\pm$0.028&2.958$\pm$1.967&0.049$\pm$0.013&7.529$\pm$4.109\\
		\hline
		\hline
		\checkmark &\checkmark&&0.887$\pm$0.264&2.644$\pm$1.469&0.048$\pm$0.014&6.158$\pm$3.539\\
		\hline
		\hline
		\checkmark&&\checkmark w/o pretrain&0.865$\pm$0.027&3.011$\pm$1.635&0.050$\pm$0.015&7.551$\pm$4.435\\		
		\checkmark&&\checkmark&\colorbox{gray!40}{0.892$\pm$0.028}&\colorbox{gray!40}{2.308$\pm$0.967}&0.049$\pm$0.016&6.248$\pm$3.577\\
	\checkmark &\checkmark&\checkmark&0.881$\pm$0.025&3.091$\pm$1.557&\colorbox{gray!40}{0.043$\pm$0.013}&\colorbox{gray!40}{5.457$\pm$3.489}\\
		\hline		
	\end{tabular}
\end{table}

\begin{figure*}[!t]
	\centering
	\includegraphics[width=1\linewidth]{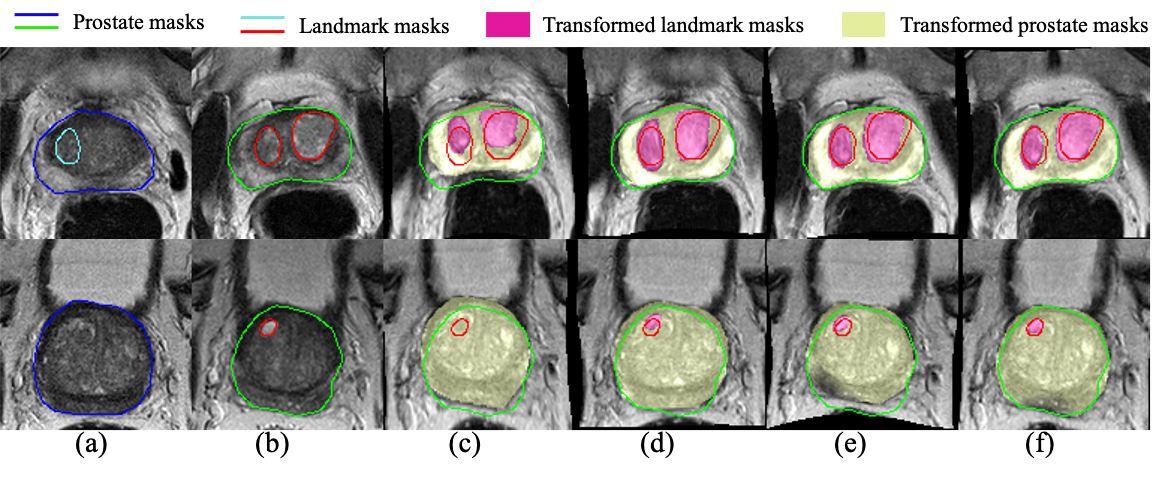}
	\caption{\textbf{Effect of proposed method.} \textbf{(a)}~moving image. \textbf{(b)}~fixed image. \textbf{(c)}~w/o quantization. \textbf{(d)}~w/ $\DV$. (e)~w/ $\DV~\text{and}~
	\DHH$. (f)~w/ $\DV, \DHH~\text{and}~\DC$.}
	\label{fig:seg} 
\end{figure*}

\noindent\textbf{Hierarchical and Collaborative Quantization.}
%
The hierarchical quatization scheme mixes the global and local information, and obtains the best results in \cref{ablation:size}.
%
We also consider randomly initializing the values of $\DC$. Its gain against the single-quantizer baseline is marginal, 
%
%
%
but once initialized by segmentation feature vectors, the collaborative embedding improves the spatial alignment to focus on the local semantic discrepancy, and obtains a better registration performance.
We provide qualitative comparison results in~\cref{fig:seg}. The proposed method performs well on aligning local patterns to the fixed image.
%
We evaluate the dictionary size of $\DV$ and $\DC$, shown in~\cref{fig:hp}, which suggests a dictionary size of 512 and 1024 for $\DV$ and $\DC$ respectively is the best option.

\noindent{\textbf{Inter-Subject Extension.}} To explore quantization for further generic application, we also validate the proposed method on inter-subject prostate MR data~\cite{bloch2015nci}. We notice that the performance increases with quantization~(DSC: $0.80\pm0.11\rightarrow0.86\pm0.04$, CD: $4.17\pm2.43\rightarrow2.12\pm1.33$, MSE: $0.04\pm0.02\rightarrow0.03\pm0.02$).
This task is challenging as the presence of prostate varies from different identities.
\begin{table}[t]
    \begin{center}
	\caption{Comparison with the-state-of-the-arts prostate registration methods. }\label{table:compare}
	\begin{tabular}{l|c|c|c|c|c}
		\hline
		\textbf{Method}&\textbf{DSC}&\textbf{CD}&\textbf{MSE}&\textbf{TRE}&\textbf{Run-time}\\
		\hline
		NiftyReg \cite{nifty}&0.270$\pm$0.304&22.869$\pm$11.761&0.041$\pm$0.019&21.147$\pm$15.841&45.76\\
		VoxelMorph\cite{2019voxelmorph}&0.763$\pm$0.081&8.842$\pm$3.156&0.053$\pm$0.015&8.833$\pm$5.147&0.69\\
		DeepTag\cite{tag2021}&0.822$\pm$0.083&7.594$\pm$2.905&0.052$\pm$0.013&7.458$\pm$4.815&1.95\\
		Contrastive\cite{liu2020contrastive} &0.856$\pm$0.117&4.973$\pm$2.407&0.054$\pm$0.018&8.2166$\pm$4.407&0.31\\
    	Basic U-Net   &0.859$\pm$0.038&4.187$\pm$2.050&0.049$\pm$0.013&7.657$\pm$4.212&0.62\\
    	VAE-like &0.865$\pm$0.029&3.623$\pm$2.189&0.045$\pm$0.019&7.626$\pm$3.948&0.72 \\
		\textbf{\ourmdoel}    &0.881$\pm$0.025&3.091$\pm$1.557&0.043$\pm$0.013&5.457$\pm$3.489&0.68\\
		\hline		
	\end{tabular}
	\end{center}
\end{table}

\subsection{Comparison with Existing Methods}
We compare \ourmdoel~with a non-optimised iterative method~\cite{nifty} and some well-known deep registration methods~\cite{2019voxelmorph,tag2021,liu2020contrastive}. To further validate the encoder-decoder structure, a common U-net and a VAE framework are implemented for prostate registration.
As shown in~\cref{table:compare}, the proposed method obtains competitive results in all metrics. Remarkably, the number of negative Jacobian determinants of our method is $0.0\pm0.0$. The consuming time is also reported.
In addition, our collaborative quantization algorithm is free from additional sub-network embedding, avoiding large memory consumption. 




\section{Conclusion}
In this paper, we proposed a collaborative quantization framework for prostate MR image registration, which was named \ourmdoel.
We introduced a hierarchical quantizer that jointly regularizes the global and local latent information to benefit the displacement prediction.
In addition, we designed a collaborative dictionary that was equipped with helpful anatomical structure knowledge to perceive the local semantic discrepancy.
The experiments showed that this method performed favorably against state-of-the-art registration methods and bypassed the overfitting problem for our dataset with a moderate size. Representing and quantizing inter-subject cues for registration can be our future work.

\subsubsection{Acknowledgements} 
This work was supported by the International Alliance for Cancer Early Detection, an alliance between Cancer Research UK [C28070/A30912; C73666/A31378], Canary Center at Stanford University, the University of Cambridge, OHSU Knight Cancer Institute, University College London and the University of Manchester. This work was also supported by the Wellcome/EPSRC Centre for Interventional and Surgical Sciences [203145Z/16/Z].


%
%
%
%
\bibliographystyle{splncs04}
\bibliography{egbib}
%
%
%
%
\end{document}